\documentclass[prb,twocolumn,showpacs,preprintnumbers,amsmath,amssymb]{revtex4}

\usepackage{epsfig}
\usepackage{graphicx}
\usepackage{dcolumn}
\usepackage{bm}

\begin{document}

\title{Edge and Surface States in the Quantum Hall Effect in Graphene}
\author{A. H. Castro Neto$^1$,   
F. Guinea$^{1,3}$, and N. M. R. Peres$^{1,2}$}

\affiliation{$^1$Department of Physics, Boston University, 590 
Commonwealth Avenue, Boston, MA 02215,USA}
\affiliation{$^2$Center of Physics and Department  of Physics,
Universidade do Minho, P-4710-057, Braga, Portugal}
\affiliation{$^3$Instituto de Ciencia de Materiales de Madrid, CSIC,
 Cantoblanco E28049 Madrid, Spain}

\begin{abstract}
We study the integer and fractional quantum Hall effect on a honeycomb lattice at half-filling (graphene) in the presence of disorder and electron-electron interactions. We show
that the interactions between the delocalized chiral edge states (generated by the magnetic field) and Anderson-localized surface states (created by the presence of zig-zag edges) lead to edge reconstruction.
As a consequence, the point contact tunneling on a graphene edge has a non-universal
tunneling exponent, and the Hall conductivity is not perfectly quantized in units of $e^2/h$. 
 We argue that the magneto-transport properties of graphene depend strongly
 on the strength of electron-electron interactions, the amount of disorder, and the details of the edges.
\end{abstract}

\maketitle

\section{Introduction}

Recent progress in development of gate and magnetic field controlled, two-dimensional (2D), graphitic devices \cite{Netal04,outros} has not only opened doors for Carbon microelectronics, but also renewed the interest in the study of strongly interacting, low dimensional, electronic systems. Graphene is a 2D Carbon material with a honeycomb lattice
and one electron per $\pi$ orbital (half-filled band), whose elementary excitations are Dirac electrons that reside at the corners of the Brillouin zone. These excitations have linear dispersion relation, $\epsilon_{\pm} ({\bf k}) = \pm v_{\rm F} |{\bf k}|$, with a characteristic Dirac-Fermi velocity
$v_{\rm F}$. All electronic properties of graphene are determined by the physics of Dirac fermions which are quite anomalous when compared to the ones found in ordinary electrons: the absence of dynamical screening \cite{mele}, a non-Fermi liquid quasi-particle lifetime \cite{GGV96}, and anomalous scattering by impurities \cite{hirsch}. Moreover, in the presence of strong disorder, graphite
samples (which are obtained from stacking of graphene layers) 
become ferromagnetic \cite{magnet,ferro} indicating the important interplay between
disorder and electron-electron interactions in these materials.
We have recently shown that because of the low dimensionality, disorder, particle-hole asymmetry, and strong Coulomb interactions, graphene presents the phenomenon of self-doping in which extended defects, such as dislocations, disclinations, edges, and micro-cracks, shift the chemical potential away from the Dirac point to produce electron or hole pockets \cite{nuno}. 
The presence of localized disorder, such as vacancies and adatoms, leads also to non-trivial
physical effects that must be understood in order to interpret the data correctly.

\begin{figure}[htb]
\centerline{\includegraphics[width=8.5cm]{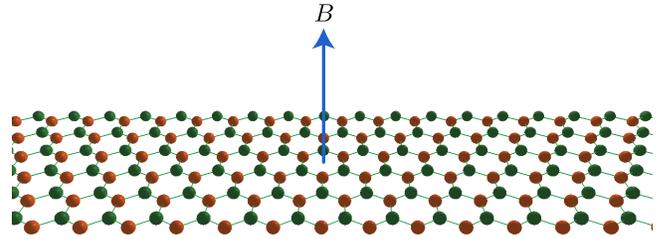}}
\caption{\label{tube} (color online) Perspective view of the Hall geometry used in our calculations
with periodic boundary conditions in one direction and 
zig-zag edges in the other. The magnetic field $B$ is applied perpendicular
to the graphene plane.}
\end{figure}

In this letter we investigate transport properties of graphene under high magnetic fields in the Hall geometry shown in Fig.~\ref{tube}. 
When a high magnetic field, $B$, is applied
to a 2D material, the electronic bulk develops Landau levels
which in the case of Dirac fermions have energy $\epsilon_n = \pm v_{\rm F} l_{\rm B}^{-1} \sqrt{n}$, where $n$ is a positive
integer, $l_{\rm B} = \sqrt{\Phi_0 / B}$ is the cyclotron radius, and 
$\Phi_0 =c h/e$ is flux quanta. Thus, the bulk of the system is gapped by the cyclotron energy scale, $\hbar \omega_c = \sqrt{2} v_{\rm F} \hbar/l_{\rm B}$ (which is much larger than the Zeeman
energy, $g \mu_B B$, where $g \approx 2$ and $\mu_B$ the Bohr magneton \cite{note3}). 

In the integer quantum Hall effect (IQHE), the bulk states are gapped and localized due to the disorder, and the electronic conduction in a Hall bar occurs through its edges \cite{halperin}.
In the case of graphene, due to the Dirac fermion nature of its
carriers, the Hall conductivity is given by \cite{nuno,sharapov}:
\begin{eqnarray}
\sigma_{{\rm IQHE}} =  (2 N +1)  \frac{2 e^2}{h}  \, ,
\label{iqhe} 
\end{eqnarray}
where $N$ is an integer, and $e$ is the electron charge.
Besides supporting bulk states a graphene Hall bar,
such as the one shown in Fig.~\ref{tube}, also supports
{\it surface} states \cite{WS00,note}. The result (\ref{iqhe})
is only valid if the surface states do not contribute to the
conduction. 

In this paper we study the integer and fractional quantum Hall
effect (FQHE) in graphene taking into account the edge and surface states of a graphene Hall bar. We show that even in the presence of disorder, when the surface states become localized, they have a direct effect in the magneto-transport. We show that the 
quantization of the conductivity, as given in (\ref{iqhe}), is not exact by the presence of surface states and becomes
dependent on the details of the sample such as the amount of disorder. Our prediction for the FQHE can be verified experimentally in graphitic devices \cite{Netal04,outros}.

The paper is organized as follows: in Section \ref{model} we present the
Hamiltonian for the problem and show how surface and edge Hall modes
originate in a graphene Hall bar; in Section \ref{edge_state} we discuss the
theory for the edge modes; Section \ref{surface} contains the theory for the
surface states in the presence of disorder and electron-electron
interactions; in Section \ref{reconstruction} we discuss the Coulomb
interaction between edge and surface states and 
the phase diagram of a graphene Hall edge in the presence of disorder 
and electron-electron interactions; Section \ref{conclusions} contains our conclusions. 
We have also included one appendix with the details of the calculations.

\section{The model}
\label{model}

The kinetic energy of electrons in graphene is described by the Hamiltonian (from now on, 
we use units such that $\hbar =1 = k_B$):
\begin{equation}
{\cal H}_{\rm kinetic} = - t \sum_{\sigma ; \langle i,j \rangle} c^\dag_{i,\sigma} c_{j, \sigma} + t'
\sum_{\sigma ; \langle \langle i,j \rangle \rangle } c^\dag_{i,\sigma} c_{j, \sigma} + h. c. \, ,
\label{hamil} 
\end{equation}
where $c_{i,\sigma}$ ($c_{i,\sigma}^{\dag}$) annihilates (creates) electrons
at the site ${\bf R}_i$ with spin $\sigma$ ($\sigma=\uparrow,\downarrow$), 
$t$ and $t'$ are the nearest neighbor and next-nearest
neighbor hopping energies, respectively. At long-wavelengths the electronic
dispersion is given by: 
\begin{eqnarray}
\epsilon_{\pm} ({\bf k}) \approx 3 t' \pm v_{\rm F} | {\bf k} | +
\frac{9 t' a^2 |{\bf k}|^2}{4} \, ,
\end{eqnarray}
 where 
 \begin{eqnarray}
 v_{\rm  F}  = \frac{3 t a}{2} \, .
 \end{eqnarray} 
In the presence of a magnetic field ${\bf B}$ the electronic hopping between sites
${\bf R}_i$ and ${\bf R}_j$ is modified via the Peierls substitution, that is, 
we rewrite the hopping as $t_{ij} \exp\{i \varphi_{ij} \}$
with $\varphi_{ij} = 2 \pi \int_{{\bf R}_i}^{{\bf R}_j} {\bf A}({\bf r}) \cdot d {\bf r}/\Phi_0$,
where ${\bf A}({\bf r})$ is the vector potential (${\bf A} = \nabla \times {\bf B}$). 

\begin{figure}[htb]
\centerline{\includegraphics[width=7.5cm,angle=-90]{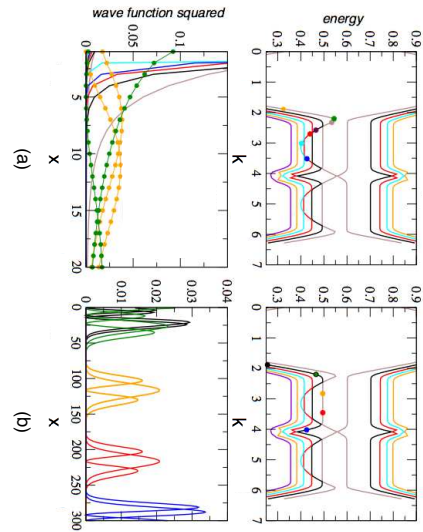}}
\caption{\label{moving} (color online) Electronic spectra (top) and wavefunctions (bottom) of a graphene Hall bar as function of the momentum $k$ parallel  to the edge. For each bullet in the top panels (going from left to right), we plot in the bottom panel the wavefunction squared as function of the distance $x$ to the edge. Note that the two top panels represent the same set of bands. (a) and (b): Surface states. (c) and (d): bulk Landau levels. Energy given in units of $t$ and distance in units of the lattice spacing $a$. }
\end{figure}

The spectra of the problem for a graphene Hall bar in
the presence of a magnetic field can be calculated exactly by
solving an eigenvalue equation \cite{gpaper}.
In Fig.~\ref{moving} we show the spectra and wavefunctions for a graphene Hall bar that is periodic along the direction parallel to the edges and $300$ lattice spacing wide, such as the one shown in Fig.~\ref{tube}, described in the tight-binding limit with $t' = 0.2 t$ in the presence of a magnetic
field such that the magnetic flux per hexagon, $\Phi$, is $\Phi=10^{-3} \Phi_0$. 

One can clearly see two types of states: bulk states that become edge states because
of the finite size of the Hall bar and a surface state localized at the edge of the sample.
In the presence of $t'$ (that breaks particle-hole symmetry) the surface mode is dispersive
with a characteristic velocity: 
\begin{eqnarray}
v_{\rm S} \sim t' a \, ,
\end{eqnarray}
while edge states have a velocity: 
\begin{eqnarray}
v_{\rm E} \sim t a \, ,
\end{eqnarray}
 where $a \approx \sqrt{3} \times1.42 \AA$ is the lattice spacing. 

\section{Edge States}
\label{edge_state}

In order to study the edge states it is conceptually simple to consider
a large graphene droplet instead of a Hall bar \cite{wen}. When a large magnetic field
${\bf B} = B {\bf z}$, 
is applied perpendicular to the graphene, a persistent current ${\bf J}$
flows along the edge of the droplet, being confined by an electric field
${\bf E}$ created by the termination of the graphene droplet. The value
of the current is given by:
\begin{eqnarray}
{\bf J} = \sigma_{xy} {\bf z} \times {\bf E} \, ,
\end{eqnarray}
where
\begin{eqnarray}
\sigma_{xy} = \nu \frac{2 e^2}{h} \, ,
\label{sxy}
\end{eqnarray}
is the Hall conductivity. Here, $\nu = 2 \pi \ell_B^2 \delta$, is the filling
fraction of the droplet, $\delta$ is the 2D electronic density (away from
half-filling). Notice, therefore, that the electrons will drift along the
edge with velocity $v = E c/B$ and, hence, the 2D electronic density along
the edge, $\rho_e(x,t) = \delta n(x)$ (where $n(x)$ is the displacement of 
the edge) obeys the equation: 
\begin{eqnarray}
\partial_t \rho_e(x,t) - v \partial_x \rho_e(x,t) = 0 \, ,
\end{eqnarray}
which describes a chiral motion (classically, $\rho_e(x,t) = \rho_e(x - vt)$).

The classical problem can be quantized in terms of the Fourier components
of the density: 
\begin{eqnarray}
\rho_k = \frac{1}{\sqrt{L}} \int dx e^{i k x} \rho_e(x) \, ,
\end{eqnarray}
where $L$ is the circumference of the edge, by canonical commutation:
 \begin{eqnarray}
\left[\rho_k,\rho_{-k'}\right] = \frac{\nu}{2 \pi} k \delta_{k,k'} \, .
\end{eqnarray} 
The Hamiltonian of the edge waves is then simply:
\begin{eqnarray}
H = \frac{2 \pi v}{\nu} \sum_{k>0} \rho_k \rho_{-k} \, .
\end{eqnarray}

The edge fermion operator can be constructed from the density operators
via a bosonic field, $\phi(x)$, such that:
\begin{eqnarray}
\rho_e(x) = \frac{1}{2 \pi} \partial_x \phi(x) \, .
\end{eqnarray}
The electron operator, $\Psi_e(x)$, is given by the Mandelstam construction:
\begin{eqnarray}
\Psi_e(x) \propto e^{\frac{i}{\nu} \phi(x)} \, ,
\end{eqnarray}
that preserves the commutation relations between the electron
and the density operators:
\begin{eqnarray}
\left[\rho_e(x),\Psi_e^{\dag}(y)\right] = \delta(x-y) \Psi_e^{\dag}(x) \, .
\end{eqnarray}
This result indicates that the operator carries charge $e$, as
required. Furthermore, it is easy to show that:
\begin{eqnarray}
\Psi_e(x) \Psi_e(y) = (-1)^{1/\nu} \Psi_e(y)\Psi_e(x) \, ,
\end{eqnarray}
and, thus, by the Pauli principle we must require that: 
\begin{eqnarray}
\nu = \frac{1}{m} \, ,
\label{laugh}
\end{eqnarray} 
where $m$ is an {\it odd integer}. The constraint (\ref{laugh}), when applied to
(\ref{sxy}), gives the quantization of Hall conductivity. 
Thus, it is clear that this construction can only describe Laughlin's main sequence.
For more complicated QHE sequences one has to use
multiple edge states \cite{nuno}. In this work we focus
on the case given in (\ref{laugh}). 

It is also easy to show that:
\begin{eqnarray}
\langle \phi(x) \phi(0) \rangle = - \nu \ln(x) + {\rm constant} \, ,
\end{eqnarray}
and, hence,
\begin{eqnarray}
\langle \Psi^{\dag}_e(x) \Psi_e(0) \rangle \propto e^{\frac{1}{\nu^2} 
\langle \phi(x) \phi(0) \rangle} \propto \frac{1}{x^m} \, ,
\label{corr}
\end{eqnarray}
as expected. In terms of path integrals, the action for the 
1D {\it chiral} Luttinger liquid action reads:
\begin{eqnarray}
S_{\rm edge} = \int_{x,t} \frac{m}{4 \pi} \left[
\partial_t \phi \partial_x \phi  - v_{\rm E} \left(\partial_x \phi(x)\right)^2
\right] \, ,
\label{chiral}
\end{eqnarray}
where $\phi(x,t)$ is a bosonic chiral field along the edge at position $x$ and time $t$. 

This construction stresses the robustness of the Hall effect: the edge state 
being chiral in nature, cannot suffer any backscattering. It is exactly the
electron-electron backscattering interaction that creates density wave and
superconducting states, and the impurity backscattering interaction that
leads to Anderson localization \cite{vojt}. Therefore, chiral edge states
iarenot influenced either by electron-electron interactions or disorder. 
Finally we note that the exponent $m$ is determined entirely by the bulk
of the system and has topological origin \cite{wen}. 

In the next section we are going to show that {\it forward} scattering
interactions between chiral edge states and surface states modify the
chiral action (\ref{chiral}) introducing instabilities in the chiral
states at finite momenta. If this is the case, it is clear that 
the relation (\ref{laugh}) (or (\ref{corr}) will not hold, spoiling the 
perfect quantization of the Hall conductivity.

\begin{figure}[htb]
\centerline{\includegraphics[width=6.5cm,angle=0]{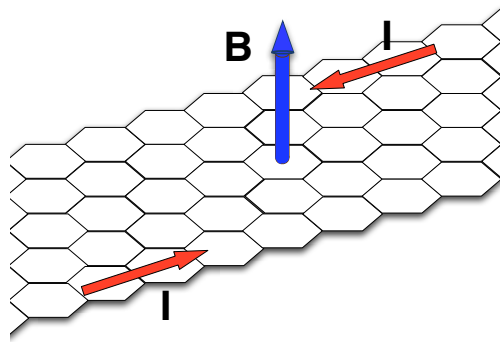}}
\caption{\label{edge} (color online) Perspective view of edge currents, $I$, on a graphene Hall bar. }
\end{figure}

\section{Surface states and disorder}
\label{surface}

The graphene surface state is also a 1D state that can be described by a {\it non-chiral} Luttinger liquid action \cite{vojt}:
\begin{eqnarray}
S_{\rm surf} = \int_{x,t} \frac{1}{2 \pi K} \left[
\frac{1}{u} \left(\partial_t \theta(x)\right)^2 - 
u \left(\partial_x \theta(x)\right)^2\right] \, ,
\label{luttinger}
\end{eqnarray}
where $\theta(x,t)$ is a bosonic field at the edge, 
\begin{eqnarray}
u = v_{\rm S} \left[1 - f^2/(4 \pi^2 v_{\rm S}^2)\right]^{1/2} \, ,
\end{eqnarray}
is the renormalized Luttinger liquid velocity 
($f$ is the electron-electron forward scattering coupling constant), and
\begin{eqnarray}
K = \left[(1-f/(2 \pi v_{\rm S}))/(1+f/(2 \pi v_{\rm S}))\right]^{1/2} \, ,
\label{lut_par}
\end{eqnarray}
is the Luttinger parameter that measures the decay of the surface correlation
functions \cite{vojt}. The surface density, $\rho_s(x)$,
is written as:
\begin{eqnarray}
\rho_s(x)  =  \frac{1}{2 \pi} \partial_x \theta(x)  \, .
\end{eqnarray}
Notice that unlike the edge mode, the surface mode is sensitive to the
electron-electron interactions and also disorder. 

\begin{figure}[htb]
\centerline{\includegraphics[width=6.5cm,angle=0]{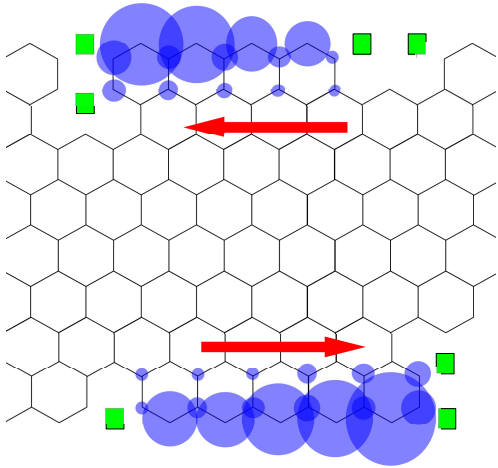}}
\caption{\label{edge_disorder} (color online) Schematic representation of a graphene Hall bar with a disordered zig-zag edge and its {\it electronic puddles} (see ref. [\onlinecite{kobayashi}]). Squares represent missing Carbon atoms, the circles' radii represent the amplitude of the localized surface electronic
wavefunction and the formation of electronic puddles. The arrows show the direction of the edge currents. }
\end{figure}

While the problem
without disorder can be easily studied theoretically \cite{chamon} (see below),
in most of this paper we focus on the realistic case of a disordered edge, such as the
one shown in Fig.~\ref{edge_disorder}. Missing Carbon atoms at the edge of a graphene hall bar effectively "cuts" the electronic wavefunction and leads to lateral confinement of the electrons. 
The lateral confinement of the electrons leads to the discretization of the surface state energy levels, as one would have for a particle moving in a 1D box. Hence, these {\it electronic puddles} have a characteristic excitation energy scale, or gap, of the order of 
\begin{eqnarray}
\Delta(\ell) \sim \frac{v_S}{\ell}  \sim \frac{t' a}{\ell} \, ,
\label{deltal}
\end{eqnarray}
 where $\ell$ is the size of the 1D domain.  
This effect is clearly seen in scanning tunneling microscopy (STM) studies of the surface of graphite \cite{kobayashi}.  

The presence of disorder at the edge of the graphene Hall bar 
leads to a back-scattering of the surface electron states which
can be written as:
\begin{eqnarray}
S_{\rm BS} = \int_{x,t} \left( V(x) e^{i \left(\sqrt{2} \theta(x,t) + 2 k_F x\right)}
+ {\rm h.c.} \right) \, ,
\label{random_p}
\end{eqnarray}
where $V(x)$ is the scattering potential (notice that only the $2 k_F$ component of the disorder potential contributes at low energies, and that the forward scattering part of the potential is irrelevant). In the presence of disorder $V(x)$ is a random variable with probability:
$
{\cal P}[V(x)] \propto \exp\left\{- \int dx |V(x)|^2/V_0^2\right\}
$
so that, after averaging over disorder:
$
\left[ V(x) V^*(y) \right]_{\rm disorder} = V_0^2 \delta(x-y) \, .
$
Hence, $V_0$ provides a measure of the amount of disorder
in the system. One possible way of dealing with (\ref{random_p})
is via a replica-trick \cite{giamarchi}. In this case, one has to add a new term to the Luttinger
liquid action (\ref{luttinger}):
\begin{eqnarray}
S_{\rm BS} = - V_0^2 \int_{x,t,t'} 
\sum_{i,j} \cos[\sqrt{2} (\theta_i(x,t)-\theta_j(x,t'))] \, ,
\label{replica}
\end{eqnarray}
where $\theta_i(x,t)$ indicates the field $\theta$ in the $i^{\rm th}$
replica. Notice that this term is highly non-local because of the quenched disorder. 
In the absence of edge modes, the full action (\ref{luttinger}) plus (\ref{replica}) can
be understood via a renormalization group (RG) calculation assuming
the disorder to be weak, that is, we define a dimensionless disorder
strength, $D$:
\begin{eqnarray} 
D \approx \frac{2 V_0^2 a^2}{\pi u^2} \, ,
\end{eqnarray}
and obtain the RG equations\cite{giamarchi} in leading order in $\delta$:
\begin{eqnarray}
\partial_{\ell} K^{-1} &=& \frac{D}{2} \, ,
\nonumber
\\
\partial_{\ell} D &=& \left(3 - 2 K\right) D \, ,
\label{rg}
\end{eqnarray}
where $\ell = \ln(W_0/W)$ is the RG scale ($W$ is the running energy
cut-off of the bosons, $W_0 \sim t'$ is the bare cut-off). It is easy
to see that disorder is irrelevant if $K>3/2$, and
it is relevant if $K<3/2$, under the RG flow.
Notice that from (\ref{lut_par}) we have $K<1$ for repulsive interactions ($f>0$) and therefore the above RG indicates that disorder always
flows to strong coupling, $D(\ell \to \infty) \to \infty$, and strong
interactions, $K(\ell \to \infty) \to 0$, as expected. This result
implies that the RG breaks down at certain scale where 
$D(\ell^*) \approx 1$ and the surface states become Anderson localized.
The localization scale can be estimated from (\ref{rg}) 
by introducing a localization length, $\xi$, so that:
$
\ell^* = \ln(W_0/(u \xi^{-1}) \, ,
$
and from (\ref{rg}) one finds:
\begin{eqnarray}
\xi \approx a \, \, D_0^{-1/(3 - 2 K^*)}
\label{loc_len}
\end{eqnarray}
where $D_0$ is the bare amount of disorder in the
system and $K^* \approx K(\ell^*)$. Notice that this
result indicates that there is a characteristic energy
scale, $E_{{\rm loc}}$, associated with the disorder
which is of the order of $E_{{\rm loc}}(\xi) \sim v_S/\xi 
\sim t' \,  D_0^{1/(3 - 2 K^*)}$. Direct comparison with
(\ref{deltal}) shows that $\xi$ can be thought as the
typical size of the electronic puddles at the edge with
a gap in the energy spectrum given by $E_{{\rm loc}}$.

This result indicates
that the bosonic correlations at larger distances decay
exponentially with $\xi$ (the time correlations are also
short ranged with a characteristic time scale $\tau_{\rm loc} \sim 
1/E_{{\rm loc}} \sim \xi/v_{\rm S}$). In this case, it is reasonable to replace
the Luttinger liquid action (\ref{luttinger}) by:
\begin{eqnarray}
S_{\rm s} \approx - \int_{x,t} \frac{u}{2 \pi K} \left[
\frac{1}{\xi^2} \theta^2(x) + 
\left(\partial_x \theta(x)\right)^2\right] \, ,
\label{dirty}
\end{eqnarray}
so that: 
$
\left[\langle \theta(x,t) \theta(0,0) \rangle\right]_{\rm disorder} \approx \delta(t) e^{-x/\xi} \, ,
$
for $x \gg \xi$ and $t \gg \tau_{\rm loc}$.

\section{Electron-electron interactions and edge reconstruction}
\label{reconstruction}

Because the edge and the surface states are confined to a small region in
space they interact with each other via a Coulomb interaction. This interaction leads
to a new term in the problem:
\begin{eqnarray}
S_{\rm coupling} = -  \frac{\lambda}{\pi} \int_{x,t} \, \, \partial_x \theta \, \, \partial_x \phi \, ,
\label{coupled}
\end{eqnarray}
where,  
\begin{eqnarray}
\lambda = \frac{e^2 \, a}{4 \pi \ell_{\rm B}} \, ,
\end{eqnarray} 
is the strength of the electron-electron coupling. 

\subsection{Clean surface states}

Let us consider first the case of a clean surface state 
interacting with a chiral edge state.
The edge state is described by the chiral Luttinger liquid 
Hamiltonian given by:
\begin{equation}
H_\chi = \sum_{q> 0} \, q v_S a^\dag_qa_q\, ,
\label{hchiral}
\end{equation}
where $a_q$ ($a^{\dag}_q$) annihilates (creates) a chiral boson
with momentum $q$, while the surface state is described by 
the Hamiltonian:
\begin{equation}
H_{LL} = \sum_{q> 0} \, q v_E (b^\dag_qb_q+b^\dag_{-q}b_{-q})
+\sum_{q> 0}\bar V(q)(b^\dag_qb_q+b^\dag_{-q}b_{-q})\, ,
\label{hll1}
\end{equation}
where $\bar V(q)$ is the surface state forward scattering interaction.
Let us consider a generic surface-edge interaction potential, $V(q)$,
and the interaction Hamiltonian:
\begin{equation}
H_{\chi,LL}= \sum_{q> 0} V(q)
[a^\dag_q(b^\dag_q+ b_{-q})+a_q(b^\dag_{-q}+ b_q)]\,.
\label{hlc1}
\end{equation}
The Hamiltonian $H_{LL}$ can by diagonalized via a 
Bogoliubov-Valatin transformation\cite{Mahan},
leading to new bosonic modes $c_q$ and $d_q$:
\begin{equation}
H_{LL}= \sum_{q> 0} E(q) (c^\dag_qc_q+d^\dag_{q}d_q)\,
\label{hll2}
\end{equation}
with $E(q)=\sqrt{q^2 v_E^2-[\bar V(q)]^2}$\,.
Under the transformation the interaction Hamiltonian becomes:
\begin{equation}
H_{\chi,LL}= \sum_{q> 0}  \tilde V(q)
[a^\dag_q(d^\dag_q+ c_q)+a_q(d_q+ c^\dag_q)]\,,
\label{hlc2}
\end{equation}
with $\tilde V(q)= V(q)[\cosh(\lambda_q)-\sinh(\lambda_q)]$,
$\cosh(2\lambda_q)=qv_F(q)/E(q)$, and $\sinh(2\lambda_q)=\bar V(q)/E(q)$.
The Hamiltonian composed by the sum of (\ref{hchiral}), 
(\ref{hll2}), and (\ref{hlc2}), is the form of the effective
Hamiltonian described in ref.~[\onlinecite{eggert96}], 
and can be diagonalized by a generalized Bogoliubov-Valatin transformation
\cite{eggert96,note1}. Introducing a spinor field
$\Psi^\dag=(a^\dag_q,c^\dag_q,d_q)$ the total Hamiltonian reads:
\begin{equation}
H= \sum_{q> 0}\Psi^\dag D \Psi - \sum_{q> 0}E(q)\,,
\label{hfinal}
\end{equation}
where $D$ in the grand-dynamical matrix\cite{colpa78}. 
The Hamiltonian (\ref{hfinal}) has the 
form (apart from constant terms):
\begin{equation}
H=\sum_{q> 0}[
\omega_\alpha(q)\alpha^\dag_q\alpha_q+
\omega_\beta(q)\beta^\dag_q\beta_q+
\omega_\gamma(q)\gamma^\dag_q\gamma_q]\,,
\end{equation}
where, after diagonalization, 
and the new
quasiparticles operators read:
\begin{equation}
\Psi^\dag (T^\dag)^{-1}=(\alpha^\dag_q,\beta^\dag_q,\gamma_q)\,,
\end{equation}
where the matrix $T$ as the form considered in
ref.~[\onlinecite{eggert96,note1}]. 

\begin{figure}[t]
\begin{center}
\includegraphics*[width=.8\columnwidth]{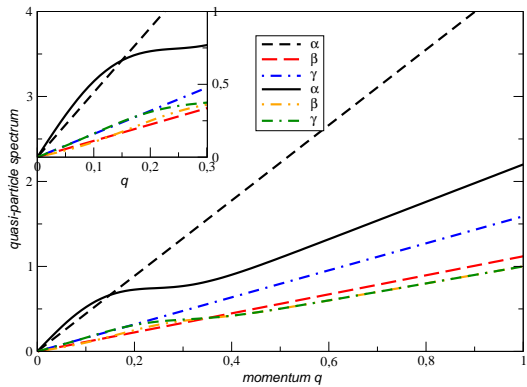}
\end{center}
\caption{(color online) Quasi-particles spectrum for the two
potentials considered in the text:
({\it i}) dashed line; ({\it ii}) solid line. The parameters are
$V_1=1.1$, $V_2=1.2$, $V_3=1.3$, $a_1=10.1$, $a_2=10.2$, $a_3=10.3$,
$v_S=2.2$, $v_E=1.$.}
\label{dispersion}
\end{figure}

The diagonalization of the Hamiltonian
(\ref{hfinal}) amounts to  find the values of the angles
$\theta$, $\phi$, and $\eta$ such that the matrix $T^\dag A T$
has non-zero diagonal elements only. All matrix elements
of matrix $A$ are given in appendix \ref{matrixA}. As in
ref.~[\onlinecite{eggert96}], 
the relation  $\omega_\alpha(q)+\omega_\beta(q)-\omega_\gamma(q)=q v_S$
holds. We have solved the eigenvalue problem for two different kinds
of electron-electron  potentials: ({\it i}) a contact potential
given by $V(x)=V_0 a \delta(x)$; ({\it ii}) $V(x)= V_0 \exp(-(|x|/a))$.
The dispersion of the bosonic modes is shown in Fig.~\ref{dispersion}.
Although both potentials are short ranged, case ({\it ii}) introduces
a momentum scale $k_s\sim 1/a$ where the spectrum deviates
significantly from the sound-like behavior obtained with 
potential ({\it i}). It is clear from these results that although
the coupling between the edge modes and surface modes alters 
the dispersion at finite wavelengths it does not lead to any
instabilities in the clean case. As we are going to show in
what follows, the presence of disorder changes this picture
significantly.

\subsection{Anderson localized surface states}

If we assume the surface mode is localized as described by
(\ref{dirty}) one can trace the surface mode completely from
the problem. In fact, using (\ref{dirty}) and (\ref{coupled})
we find that the surface mode is {\it pinned} by the edge mode:
\begin{eqnarray}
\theta(x,t) \approx - \lambda \xi^2 K u^{-1} \partial_x^2 \phi(x,t) 
\end{eqnarray}
and therefore, $\rho_s(x,t) \propto  \partial^2_x \rho_e(x,t)$. 
The pinning of the surface mode by the edge mode has a rather
interesting physical interpretation: in the presence of electron-electron
interactions the surface mode is {\it dragged}
by the edge mode in its motion. The dragging described here has
similarities with the Coulomb drag between coupled {\it clean} non-chiral Luttinger
liquids in quantum wires \cite{nazarov,klesse,minnesota} but differs
from it in some fundamental ways: ({\it i}) the chiral edge state is
a persistent current, is not subject to backscattering, and hence
can only interact through forward scattering (small momentum transfer), 
as described in (\ref{coupled}); ({\it ii}) the non-chiral Luttinger liquid is
localized by impurities and hence electrons are not free to move (unless
the applied force by the chiral component 
is greater than a threshold that leads to the depinning
of the localized state, a situation not considered here) .  
Hence, although there is no macroscopic voltage drop along the edge,
the force applied by the edge over the surface state will lead
to microscopic voltage drops (charge accumulations in the electronic
puddles, see Fig.~\ref{edge_disorder}). The final picture can be
summarized in terms of the scattering of the edge electrons by 
the potential created by the surface states.  

Since the theory described by (\ref{chiral}), (\ref{dirty}), and (\ref{coupled}) is gaussian, 
the surface states can be exactly
traced out of the problem. The effective action for the chiral
modes then reads:
\begin{eqnarray}
S = \frac{m}{4 \pi} \int_{k,\omega}
k \left[\omega  - v_{\rm E} k \left(1 - g \frac{k^2}{k^2+\xi^{-2}} \right)\right]
|\phi(k,\omega)|^2 \, ,
\end{eqnarray}
where, 
\begin{eqnarray}
g = \frac{2 \lambda^2 K}{m u v_E} \, ,
\end{eqnarray}
is the surface-edge coupling. 
Notice that the dispersion of the chiral modes is given by:
\begin{eqnarray}
\omega(k)  = v_{\rm E} k \left(1 - g \frac{k^2}{k^2+\xi^{-2}} \right) \, ,
\label{disp}
\end{eqnarray}
and, hence, for $k \ll \xi^{-1}$ the chiral mode dispersion
becomes: $\omega_k \approx v_{\rm E} (k -\kappa k^3)$
where $\kappa = g \xi^2$, and for $k \gg \xi^{-1}$
one finds: $\omega_{k} \approx v_{\rm E}[1-g] k$.

\begin{figure}[htb]
\centerline{\includegraphics[width=6.5cm,angle=0]{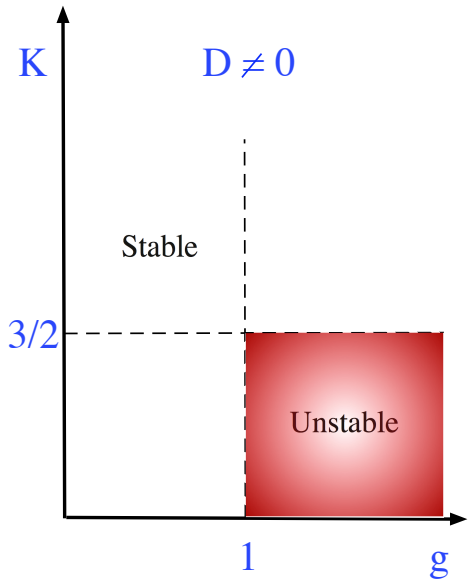}}
\caption{\label{phase_diagram} (color online) Phase diagram of the problem,
in the presence of disorder $D_0 \neq 0$,
as a function of the Luttinger liquid parameter, $K$, and 
surface-edge coupling, $g$. }
\end{figure}

At long wavelengths ($k \ll \xi^{-1}$), that is, 
distances larger than the localization length, the surface mode
does not affect the edge mode.
The fact that the dispersion at short wavelengths can 
become negative if $g>1$ indicates the existence of
an instability (a quantum critical point) at finite wavevectors. 
It is easy to see that the dispersion (\ref{disp}) vanishes at
$k=k_c$ where: 
\begin{eqnarray}
k^{-1}_c \sim \xi \sqrt{g-1} \, ,
\label{kc}
\end{eqnarray} 
for $g >1$. Therefore, the spectrum of the edge
mode becomes negative for $k > k_c$ indicating that the edge state
becomes unstable. Notice that while disorder (and hence, $\xi$)
determines the length scale $k_c^{-1}$ of the instability, this
instability only occurs for a value of $g$ above a critical value $g_c =1$ which marks a quantum phase transition in the
problem. For $g<g_c$ the edge mode is stable but  
for $g>g_c$ and for {\it any} amount of
disorder there is an instability in the system with characteristic
length scale given by (\ref{kc}). The phase diagram of the edge mode as a function
of the Luttinger parameter, $K$, and surface-edge coupling, $g$, has, therefore, 
the structure shown in Fig.~\ref{phase_diagram}.

In order to numerically estimate
the experimental value of this instability, let us consider
the case of a weakly interacting surface state ($K\approx 1$, $u\approx v_{\rm S}$) so that:
\begin{eqnarray}
g \approx \frac{e^4}{t t'} \delta \, .
\end{eqnarray}
Hence, there is a critical density $\delta_c$ such that $g = g_c=1$ given by:
\begin{eqnarray}
\delta_c \approx \frac{t t'}{e^4} \, ,
\end{eqnarray} 
so that $k^{-1}_c \to \infty$ at this point even for small amount of disorder.
Clearly, in the absence of disorder ($D_0 =0$, see (\ref{loc_len}) ) $\xi \to \infty$ we have $k_c =0$ at the outset and the instability cannot occur.  
Notice that $\delta_c$ is
independent of the disorder, depending only on the ratio between
kinetic to Coulomb energies in the system. 
Although there is uncertainty\cite{com} on the value of $t'$,
if we use $t' \approx 0.1 t \approx 0.2$ eV, $e^2 \approx 16$ eV \AA,  
one finds $\delta_c \approx 10^{12} - 10^{13}$ cm$^{-2}$, which is the order of magnitude of carriers
in these materials \cite{Netal04,outros}. 

As shown in Ref.~[\onlinecite{yang}], this instability
is an indication of a quantum Hall edge reconstruction where
new low energy modes are generated at the edge. Edge
reconstruction has been proposed to be important for the
understanding of IQHE \cite{chamon,recoiqhe} as well as FQHE \cite{recofqhe}
in semiconducting devices
and for the interpretation of point contact tunneling between
a Fermi liquid and a quantum Hall edge. In fact, the current-voltage
characteristics for point contact tunneling follows a power law,
$I \propto V^{\alpha}$, where the exponent $\alpha$, in the
absence of reconstruction, is supposed to be universal and
independent of the details of the edge. Nevertheless, recent 
experiments show a different picture \cite{exp}. The discrepancy
between theory and experiment can be assigned to edge
reconstruction. We expect a similar effect to occur in graphene
and graphite.

\section{Conclusions}
\label{conclusions}

The dragging of the surface mode by the edge mode has also
consequences for the magneto-transport. The longitudinal, $\rho_{\rm xx}$,
and Hall, $\rho_{\rm xy}$, resistivities depend directly on whether the
electronic states are localized or not. When the electronic states are
localized and the longitudinal conductivity, $\sigma_{\rm xx}$, vanishes,
one has $\rho_{\rm xy} = 1/\sigma_{\rm QHE}$ and $\rho_{\rm xx} =0$
and therefore a perfect quantization of the Hall resistivity in units of 
$h/e^2$.  

We have shown that the presence of edge disorder affects directly
the physics of the edge states allowing for the possibility of edge
reconstruction. The effect is stronger in the FQHE than in the IQHE
because of the characteristic energy scales in the problem. In the IQHE the dominant energy scale is the cyclotron energy, $\omega_c$. 
When $\omega_c$ is larger than the broadening of the Landau levels due to disorder, the IQHE becomes observable. 
For a magnetic field $B \approx 6$ T we have $\omega_c \approx 1,000$ K,  which is much larger than the cyclotron energy in conventional semiconducting Hall bars (which is of order of a few K).
This rather large cyclotron energy (a result of the Dirac dispersion) makes the observation of the IQHE relatively simple \cite{Netal04,outros}. 

The situation with the FQHE is very different. 
For the FQHE what matters is the {\it bulk} energy of interaction between the electrons \cite{laughlin} which 
is of order of $e^2/ (\epsilon_0 \ell_B) \sim [e^2/(\sqrt{2} \epsilon_0 v_F)] \omega_c \sim \omega_c/\epsilon_0$ where $\epsilon_0$  is the dielectric constant of graphene (we have used that $e^2/(\sqrt{2} v_F) \sim 1$, see ref. [\onlinecite{ferro}]). Because of the presence a back
gate in the experiments, we expect screening to be as strong as in ordinary semiconducting
devices where $\epsilon_0 \approx 10$ and hence $e^2/(\epsilon_0 \ell_B) \ll \omega_c$. 
The surface mode localization, and the formation of electronic puddles, lead to changes in the local electrostatic potentials in the Hall bar and affect screening, modifying
the bulk electron-electron interactions if the samples are not wide enough (which is the
case of the current experiments where the samples are of order of $10 \mu$ m wide \cite{Netal04,outros}). Therefore, the bulk states of the FQHE will be directly affected
by surface state localization leading to a change of the longitudinal conductivity in the system. 

In the presence of a surface state, the longitudinal 
conductivity can be small but finite ($\sigma_{\rm xx} \ll \sigma_{\rm QHE}$) 
and strongly dependent on the amount of disorder 
at the edge. In this case resistivities are given by:
\begin{eqnarray}
\rho_{\rm xx} &\approx& \rho_{\rm xx}^0 \left(\sigma_{\rm xx}/\sigma_{\rm QHE}\right)^2 \, ,
\nonumber
\\
\rho_{\rm xy} &\approx& \sigma^{-1}_{\rm QHE} \left[ 1 - \left(\sigma_{\rm xx}/\sigma_{\rm QHE}\right)^2 \right] \, ,
\end{eqnarray}
where $\rho_{\rm xx}^0 = 1/\sigma_{\rm xx}$. Notice that although there is a large
reduction in the longitudinal resistivity ($\rho_{\rm xx} \ll \rho^0_{\rm xx}$ since
$\sigma_{\rm xx} \ll \sigma_{\rm QHE}$) it is still finite in the "quantum Hall regime".
At the same time, the quantization of the Hall resistivity is only partial since it will
be spoiled by a factor $(\sigma_{\rm xx}/\sigma_{\rm QHE})^2 \ll 1$. This type
of effect has been observed in graphite \cite{graphite} and we expect it to occur
in disordered graphene samples.  

In summary, we have studied the integer and fractional quantum Hall effect in
graphene taking into account edge and surface modes. We show that although
the surface modes are localized by disorder in the absence of a magnetic field,
they become delocalized by the edge modes that drag the surface modes in
their motion via electron-electron interactions. Our results indicate that in this
case the Hall edge undergoes a reconstruction leading to a non-universal
point contact exponent that depends strongly on the amount of disorder in the
system. Furthermore, we also show that a perfect Hall effect is not possible
in disordered graphene samples due to the presence of surface modes. Our results
show that the Hall resistivity is not quantized and can change significantly from
sample to sample depending on disorder and electron-electron interactions.

N. M. R. P. and F. G. acknowledge 
the Quantum Condensed Matter Visitor's Program at Boston University for
support. N. M. R. P. acknowledges Funda\c{c}\~ao para
a Ci\^encia e Tecnologia for a sabbatical grant.
A. H. C. N. was supported by the NSF grant DMR-0343790.
We thank A. Geim and T. Martin for illuminating discussions.

\appendix

\section{Matrix elements of $A$}

\label{matrixA}

The matrix $A$ has the following  diagonal  and off-diagonal elements:
\begin{widetext}
\begin{equation}
A(1,1)=
q v_S \cos ^2\phi  \cosh^2\theta -2 V(q) \cos \phi  (\sin \phi +\cos \phi  \sinh
\theta ) 
\cosh \theta +E(q) \left(\sin ^2\phi +\cos ^2\phi  \sinh ^2\theta
   \right) \, ,
\end{equation}

\begin{flalign}
A(2,2)&= \left(q v_S \cosh ^2\theta  \sin ^2\phi +\tilde V(q) \cosh \theta  \left(\sin (2\phi )
-2 \sin ^2\phi  \sinh \theta \right)+E(q) \left(\cos ^2\phi +\sin ^2\phi  \sinh
   ^2\theta \right)\right) \cosh ^2\eta &\nonumber\\
&+ \sinh(2\eta)  ((\tilde V(q) \cos \phi 
+(q v_S+E(q)) \cosh \theta  \sin \phi ) \sinh \theta -\tilde V(q) \cosh (2 \theta ) \sin \phi )
   &\nonumber\\
&+\sinh ^2\eta  \left(E(q) \cosh ^2\theta -2 \tilde V(q) \sinh \theta  
\cosh \theta +q v_S \sinh ^2\theta \right) \, ,
&
\end{flalign}

\begin{equation}
A(3,3)=
q v_S \cos ^2\phi  \cosh ^2\theta -2 \tilde V(q) \cos \phi  (\sin \phi +\cos \phi  \sinh
\theta ) 
\cosh \theta +E(q) \left(\sin ^2\phi +\cos ^2\phi  \sinh ^2\theta\right) \, ,
\end{equation}

\begin{flalign}
A(1,2)&= \tilde V(q) \cos (2 \phi ) \cosh \eta  \cosh \theta +\sinh \eta
(((q v_S+E(q)) \cos\phi  
\cosh \theta -\tilde V(q) \sin \phi ) \sinh \theta -\tilde V(q) \cos \phi  \cosh (2 \theta
   ))\nonumber&\\
&+
\frac{1}{4} \cosh \eta  \sin (2 \phi ) (q v_S-3 E(q)+(q\tilde v_S+E(q)) \cosh (2 \theta )
-2 \tilde V(q) \sinh (2 \theta ))\, ,  &
\end{flalign}

\begin{flalign}
A(1,3)&=-\tilde V(q) \cos (2 \phi ) \cosh \theta  \sinh \eta -\frac{1}{4} \sin (2 \phi ) 
(q v_S-3 E(q)+(q v_S + E(q)) \cosh (2 \theta )
&\nonumber\\
&-2 \tilde V(q) \sinh (2 \theta )) \sinh \eta +\cosh \eta  
(\tilde V(q) (\cos\phi  \cosh (2 \theta )+\sin \phi  \sinh \theta )-(q v_S + E(q)) \cos \phi  
\cosh \theta  \sinh \theta ) \, , &
\end{flalign}

\begin{flalign}
A(2,3)&=\frac{1}{8} \sinh (2 \eta ) \left(2 (q v_S - 3 E(q)) \cos ^2\phi+(q v_S+E(q)) 
(\cos (2 \phi )-3) \cosh (2 \theta )-4 \tilde V(q) \cosh \theta  \sin (2
\phi) \right. 
&\nonumber\\
&-2 \left. \tilde V(q) (\cos (2 \phi )-3)
   \sinh (2 \theta )\right)+\frac{1}{2} \cosh (2 \eta ) (2 \tilde V(q) \cosh (2 \theta ) \sin \phi 
&\nonumber\\
&-(q v_S+E(q)) \sinh (2 \theta ) \sin \phi -2 \tilde V(q) \cos \phi  \sinh \theta ) \, .&
\end{flalign}
\end{widetext}


\begin{thebibliography}{99}
\bibitem{note3}For $B = 12$ T we find $\hbar \omega_c \approx 0.142$ eV
($\approx 1651$ K) and $g \mu_B B \approx 7 \times 10^{-4}$ eV ($\approx 8$ K).
\bibitem{note}Notice that strictly speaking there is no
surface in a finite graphene layer but a 1D edge. 
In order to make distinction between QHE edge 
states generated by a magnetic field 
and graphene edge states we call the latter
surface states.
\bibitem{note1}
The matrix element $T_{33} $of the transformation $T$ given in
Ref. [\onlinecite{eggert96}] should read
$\sin\phi\sinh\theta\sinh\eta+\cosh\theta\cosh\eta$.
\end{thebibliography}
\end{document}